\documentclass[12pt,aps,nofootinbib]{revtex4}
\usepackage{amsmath}
\usepackage{graphicx}

\begin{document}

\title{ Effective field theory and  cold Fermi gases near  unitary limit.}
\author{Boris Krippa$^{1}$}
\affiliation{$^1$ School of Physics and Astronomy,
The University of Manchester, M13 9PL, UK}
\date{\today}
\begin{abstract}
Ultracold Fermi-gases near the unitary limit are studied in the framework of Effective Field Theory.
It is shown that, while one can obtain a reasonable description of the universal proportionality constants both in the   
narrow and the broad Feshbach resonance limits, the requirement of the reparametrisation invariance leads to appearance 
of the three body forces needed to cancel the otherwise arising off-shell uncertainties. The size of the uncertainties is estimated.

\end{abstract}
\maketitle
Ultracold Fermi gases have recently attracted a lot of attention ( see \cite{Rad} and references therein) due to exiting possibility of
 tuning the strength of the fermion-fermion interaction through the Feshbach resonances so that the scattering length may become much larger
then the typical scale in the system. This large scattering length is the main dynamical factor resulting in establishing the so called unitary limit
(UL) which is believed to be universal in a sense that the only relevant energy scale is given by that of the non-interacting Fermi gas

\begin{equation}
E_{GS}=\xi E_{FG}=\xi \frac{3}{5}\frac{k^{2}_{F}}{2 M},
\end{equation}
where $M$ and $k_F$ are the fermion  mass and Fermi momentum correspondingly and   $\xi$ is the universal
 proportionality constant, which does not depend on the details of the interaction.
 The other dimensional characteristics of the cold Fermi-gas in the UL such as paring energy  or chemical potential 
can also be represented in the same  way.

 Large scattering length implies nonperturbative treatment. The most ``direct''
nonperturbative method is based on the fixed-node MC approach \cite{Ca, As}. However, being potentially the most powerful calculational tool,
 direct numerical simulations still have many limitations related to finite size effects, discretization errors, trial wave function dependence
 etc which may even become amplified in certain physical situations (the system of several fermion species is one possible example). 
All that makes the development of the analytic 
approaches indispensable. Several such   approaches have been  suggested so far ranging from 
the exact renormalisation group \cite{Kr1, Di} and expansion in terms of dimensionality of space \cite{So}
to more phenomenological approaches using the density functional method \cite{Bu} and 
many body variational formalism \cite{Ha} . The ``world average'' for the value of $\xi$ is $0.42 \pm 0.002$.
It is important to emphasise that UL refers to the idealised situation with an infinite scattering length and vanishing effective 
radius which is the case of a broad Feshbach resonance.   Even small but finite effective radius sets the other scale so that the system 
may deviate from the strict UL. This is the case in the narrow  Feschbach resonance limit and also in nuclear/neutron matter where the
experimental value of 
the effective radius is only one order of magnitude smaller then the scattering length and should therefore be taken into account. The purpose
 of this paper is to analyse the system of the cold Fermi atoms in the both narrow and broad  Feschbach resonance limits in the framework of 
effective field theory (EFT) and study the general constraints EFT imposes on theoretical approaches describing
the system of cold Fermi atoms in both  UL and nearby.

EFT is based on the fact that the low scale dynamics is only weakly dependent on the details of the interaction at small distances. The low 
scale phenomena can then be described by a local effective lagrangian with some effective coupling constants reflecting the short range dynamics 
in some effective, indirect way. The physical amplitudes which can be derived from this lagrangian take the general form of the expansion in powers
of the low scales involved with implicit assumption than all low scale are ``natural'' in a sense that all the possible dimensionless ratios are
of order unity. As we have already mentioned the main feature of the   cold Fermi atoms in UL is the large value of the scattering length.
It makes the use of the canonical EFT impossible as the large scattering length introduces a new ``unnatural'' scale to the problem so that
 the power expansion is no longer valid. To overcome this difficulty it was suggested to iterate the leading term of the interaction to all orders by solving 
the  Lippmann - Schwinger (LS) equation and to treat 
the rest as a perturbation \cite{We, KSW}.

Being a proper field theory EFT must be regularised and renormalised. Besides,  it must satisfy the reparametrisation invariance requirement which means that,
although one can choose different representations for the field operators in the effective Lagrangian, the physical amplitudes should remain the same in any
 representation. In formal field theory this statement is known as a equivalence theorem \cite{He}. In a more phenomenological language it means
that the on-shell observables must be independent on the parametrisation used for the off-shell part of the fermion-fermion interaction. In the context of
 many fermion systems
 with arbitrary scattering length the physical consequences of the reparametrisation invariance were considered in \cite{Kr2}.
 In this paper we use the findings of \cite{Kr2} to analyse the system 
of cold Fermi atoms at and around UL.  The other important general requirement, usually called renormalisation group (RG) invariance, is the
 independence of the on-shell physics on the renormalisation
 parameters like cutoff or subtraction point.
 To comply with RG invariance the fermion-fermion scattering amplitude must satisfy the RG equation.

According to EFT the physical amplitudes at low scale can be derived from the effective Lagrangian with purely short range interactions. The corresponding 
 LS  equation for the fully off-shell amplitude $T(k',k,p)$   takes the form

\begin{equation}
T(k',k,p) = V(k',k,p) + M \int \frac{dq q^2}{2 \pi^2} \, V(k',q,p) 
\frac{T(q,k,p)}{p^2 - q^2 + i\epsilon}.
\label{eq:LSE2}
\end{equation}

Here we use $k$ and $k'$ to denote relative momenta and the energy dependence is given by $p = \sqrt{M E}$, the on-shell momentum 
corresponding to the centre-of-mass energy $E$.
One possible form of the interaction can be written as 
\begin{equation}
V = V_1= C_0 +   C'_2 p^2 .
\label{eq:lag1}
\end{equation}
Since it depends on on-shell momentum the interaction is purely energy dependent.  The interaction can be written in a separable form so that the LS equation can easily
be solved analytically. The resulting $T$-matrix takes the form
\begin{equation}
\frac{1}{T(p)}=\frac{(C_2 I_3 -1)^2}{C_0 + C_2^2 I_5
 + {k^2} C_2 (2 - C_2 I_3)} - I(p),
\label{eq:Tonexp}
\end{equation}
where the loop integrals are

\begin{equation}
I_n  \equiv -\frac{M}{(2 \pi)^2} \int dq q^{n-1},
\label{In}
\end{equation}
 
and  

\begin{equation}
I(p) \equiv \frac{M}{2 \pi^{2}}\int dq  \, \frac{q^2}
{{p^2}- {q^2}}.\label{eq:IEdef}
\end{equation}
These loop integrals are divergent and therefore the procedure of 
regularisation and renormalisation must be carried out. As a side remark we note that the
 issue of the nonperturbative renormalisation is quite a subtle problem.
In contrast to the standard perturbative case where the usual field
 theoretical methods can be used to regularise the given divergent graphs and then  
renormalize the bare coupling constants, in the nonperturbative situation
the renormalisation of the whole integral equation must be carried out.
In the case when the analytic solution for the scattering amplitude can be obtained,
as is the case here, the renormalisation of the
 amplitude is a rather straightforward procedure. However, if the explicit solution
 is not possible (if the long range forces are added, for example) then the special
 care is needed to perform the renormalisation in a consistent way.
In this paper we follow the procedure used in Ref. \cite {Ge} to renormalize the
effective fermion-fermion amplitude in vacuum. 
 We subtract the divergent integrals at some kinematical point $p^2 = -\mu^2$ so that all the couplings should now depend on
 the subtraction point $\mu$ to ensure that the scattering amplitude is $\mu$ independent. The regularised fermion-fermion amplitude
has the form

\begin{equation}
T_1=\frac{C_{0} + p^2 C'_{2}}{1 +\frac{M}{4\pi}(C_{0} + p^2 C'_{2})(ip + \mu)}  .
\label{eq:T1}
\end{equation}
The same expressions can be obtained in the regularisation  scheme considered in \cite{KSW}.

The coupling constants can be determined 
from the suitable observables. For example, in the  renormalisation scheme adopted in \cite{KSW} they can be related to the low energy observables as 
\begin{equation}
C_0(\mu) = \frac{4\pi}{M}\left(\frac{1}{-\mu + 1/a}\right), \qquad    C'_2(\mu)=\frac{M}{4\pi} C^{2}_0(\mu) \frac{r}{2},
\end{equation}
where $a$ and $r$ are the scattering length and  the effective radius correspondingly.
Strictly speaking the expression for the $V_1$ is written up to next-to-leading order term in the small scale according to the counting scheme developed
in \cite{KSW}. The LO coupling $C_0$ scales as $p^{-1}$ and should therefore be treated nonperturbatively. The rest can  be interpreted
 as a perturbation. However, it is rather easy to solve the whole Lippmann - Schwinger equation and obtain the vacuum $T$-matrix. It is worth mentioning that 
in the simplest case of the pointlike interactions all EFT does is just complicated way of getting the well known phenomenological 
effective range expansion. The full strength of the EFT can be easily realised when considering more complicated situations like, for example, the interaction with 
external currents.  The straightforward generalisation of the phenomenological approaches  may lead
to the conflict with gauge invariance. The EFT solves this problem in a natural way treating all the interaction terms on an equal footing. That's the main motivation 
of using more general approach even in the case when it can be reduced to the well known phenomenological approaches.

It is clear that the above written purely energy dependent form of interaction is not unique.
There may exist more general form of interaction in which the energy and momentum can be treated as formally independent variables. 
It has the form 
\begin{equation}
 V = V_2 = C_0 + C'_{2} p^2 + \frac{1}{2}C_2 ({\bf k}^2+{\bf k}^{\prime 2}-2p^2),
\label{eq:lag2}
\end{equation}
where ${\bf k}$ and ${\bf k}'$ denote the initial and final relative momenta of the fermions and the coupling $C_2$ describes a
purely off-shell interaction. The solution of the  Lippmann - Schwinger equation is again straightforward and we get
\begin{eqnarray}
T_2&=&T_1\biggl[1+\frac{1}{2(C_{0} + p^2 C'_{2})}\biggl(
C_2({\bf k}^2+{\bf k}^{\prime 2}-2p^2)\cr
&&\qquad-\frac{M}{8\pi}
C_2^2(p^2-{\bf k}^2)(p^2-{\bf k}^{\prime 2})(ip + \mu)
\biggr)\biggr],
\label{eq:T2}
\end{eqnarray}
where $T_1$ is given by Eq.~(\ref{eq:T1}). 

One can see from this equation that both amplitudes coincide on-shell so that the reparametrisation invariance is fulfilled.

The situation becomes much more complicated in the presence of medium. The corresponding amplitude can be calculated by solving the 
Feynmann-Galitskii equation
\begin{equation}
T^m = V + VG^{F}G^{F}T^m,
\label{eq:FG}
\end{equation}
where $G^F$ is  the in-medium one-fermion propagator 
\begin{equation}
G^{F}(\tilde{k}) = \frac{\theta( k - p_F)}{k_0 - \omega_k + i \epsilon} + \frac{\theta(p_F - k)}{k_0 - \omega_k - i \epsilon},
\end{equation}
here $\tilde{k} \equiv (k_0, {\bf k})$ and $\omega_k \equiv k^2 / 2 M$.

 For the interaction $V_1$ the solution of this equation takes the form
\begin{eqnarray}
T^{m}_{1}=\biggl[\frac{1}{C_{0}(\mu) + p^{2}C'_{2}(\mu)} +\frac{M \mu}{4\pi} - \frac{M (p_F + Q/2)}{4\pi^2}+ 
\frac{M}{4\pi^2}\biggl( p\log\frac{p_F + Q/2 + p}{p_F + Q/2 - p} +\nonumber\\ 
\frac{Q^{2}/4 - p^{2}_{F} +p^2}{\pi Q}\log\frac{(p_F + Q/2)^2 + p^2}{p^{2}_{F} - Q^2/4 - p^2}\biggl)\biggr]^{-1},
\end{eqnarray}
where $p_F$ is the Fermi momentum and the amplitude is written for the case of the non-zero total momentum $Q$ of the interacting fermion pair.
Having determined the in-medium fermion-fermion amplitude we can extract the universal constant $\xi$ by computing the energy density $E_{GS}$
of the interacting Fermi gas and using the Eq.(1).  The expression for the  $E_{GS}$ is given by 
\begin{equation}
E_{GS}= E_{FG} + E_{int},
\end{equation}
where 
\begin{eqnarray}
E_{int} = \frac{3}{2\pi^2 p^{5}_{F}}\biggl[\int^{2p_F}_{0} Q^2 dQ \int^{p_F-Q/2}_{0} p^2 dp T^{m}(p^2,Q) +\nonumber\\ 
\int^{2p_F}_{0} Q d Q \int_{p_F-Q/2}^{\sqrt{p^{2}_{F} - Q^{2}/4}} p dp T^{m}(p^2,Q)(p^{2}_{F} - p^2 - Q^{2}/4)\biggr]
\end{eqnarray}
 Calculation of  the universal constant $\xi$
in the genuine UL does not involve energy dependent part of the fermion-fermion interaction so we can drop the second term in Eq.(3).
 We obtain 
$\xi(UL) \simeq 0.33$ in the unitary limit with vanishing effective radius. Similar results
were obtained in \cite{Hl} using the effective range expansion for  the fermion-fermion scattering amplitude in vacuum.
In the language of EFT it corresponds to a particular choice of the subtraction point $\mu = 0$ so  for the lowest-order effective coupling we get
$C_0(\mu = 0) = 4\pi a/ M$. We see that the effective coupling becomes arbitrary large in the UL so that it is hard to
extract the effective couplings from observables and to justify the EFT expansion for the effective
 lagrangian in this case. As shown in \cite{KSW} each individual graph in the sum of the bubble diagrams for the  fermion-fermion scattering amplitude
goes as $(4\pi a/ M)(i a p)^L$, where $L$ is the number of loops so that each contribution in the bubble sum is bigger then the preceding one. 
It means that there is no well defined expansion parameter and it may result in artificial dependence on short range physics, unacceptable situation for
consistent EFT. We emphasise, however that there is nothing wrong in using the phenomenological expressions like effective range expansion for the 
fermion-fermion  amplitude to calculate the values for some physical quantities like $\xi(UL)$. In fact, a proper use of EFT leads precisely to this. 
Therefore, there is no wonder that EFT and phenomenological description lead 
to the similar results in this (simplest) case. One only needs to keeps in mind the constraints and limitations of the phenomenological approach. 
One side remark is that EFT provides a natural way of incorporating gauged interactions and external currents in unambiguous was, the opportunity often
missing in phenomenological approaches where the conflict with gauge invariance is rather rule then exception. It is clear that EFT and phenomenological approaches 
are very different on this level of complexity. Therefore, it looks more advantageous to use EFT even in the simple cases where a suitable analytic parametrization 
of the fermion-fermion amplitude like effective range expansion can formally provide similar answers.

It is important to emphasise that the particle-particle (and hole-hole) summation with undressed propagators represented by the FG equation
 provides only the simplest many-body approximation to the many-body problem. The further complications come from the self-energy and vertex corrections
involving particle-hole pairs. However, they should start contributing at the order, prescribed by the power counting. It seems possible that for the dilute systems 
the counting should not be much different from that suggested in \cite{KSW} for the fermion-fermion amplitude in vacuum since each loop involving the
at least one hole line leads to the contribution proportional to $p_F$ which is small so that the leading contribution is given by the summed 
particle-particle ladder in accord with the power counting from  \cite{KSW}. It is then looks conceivable that self-energy and vertex corrections
will contribute at higher order. There is however a subtlety here. Let's consider the Eq. (13) for the elementary fermion-fermion amplitude
and take for simplicity the case with zero total momentum. The corresponding expression for the $T$ matrix can be written as  

\begin{equation}
T^m_1 = \frac{1}{\frac{1}{T_1}
 + \frac{M}{4\pi^2}[ p\log\frac{p + p_F}{p - p_F}
 - 2p_F]}  ,
\label{eq:Tm1}
\end{equation}
In the strict UL case the term with $\log$ does not contribute and the amplitude scale as $1/p_{F}$. Being combined with the loop  it leads
to the contribution of order one. It means that particle-particle and hole-hole ladders as well as particle-hole rings should in principle be treated 
on the same footing. Formulation of the power counting prescription is an open problem in this case and the use 
of the either simplest approximation like FG equation or more complicated approaches with self-energy and vertex corrections  equally
 requires a further justification. If the in-medium power counting issue is resolved then all the improvements and 
corrections can be taken into account in a systematic way. 

The situation seems to be somewhat different for the system with the finite effective range ($p_{F}r \sim 1$)
 and nonnegligible shape parameter (see below). In such a system    these effects  may lead to some additional suppression for the loop diagrams involving 
holes so that in this case counting indeed is
 closer to the vacuum one and the corresponding self-energy and vertex corrections are of higher order and thus suppressed. In this case 
the use of the FG equation seems more justified. Of course,the relative contributions of the suppressed and unsuppressed terms depend on 
a concrete value assumed for the effective radius. It is important however to keep the values of the effective range and shape parameter within the limits
of applicability of the effective range expansion. We stress that these arguments are rather crude and a consistent power counting is yet to be formulated. 
For example, it is not quite clear what kind of diagrams should contribute at the next-to-leading order and at what value of  $p_{F}r$ should the counting start 
deviating from the vacuum one. All these issues need to be clarified in the future EFT-based studies of strongly interacting Fermi systems 
(dilute atomic Fermi gases,
neutron/nuclear matter etc). We also note that the available  experimental data on cold fermionic atoms correspond to the case of the broad 
Feshbach resonance with  $p_{F}r <<1$ so that in this context the system of cold fermionic atoms with a large scattering length and 
the finite and modest effective radius refers to some future experiments. On the other hand the many-fermion system of this type is realised in 
nuclear/neutron matter.   

 With the effective range effects  included we have obtained the value $\xi$ = 0.48  at  $p_{F}r \sim 1$.
The result is rather  close to the ``world average''
and seems to bring a certain credibility to the approach developed here. The results of calculations in the case of narrow
Feshbach resonance with finite effective radius are shown on Fig.1 as a function of the dimensionless parameter $p_F r$. One notes, that 
the contribution due to the  pairing interactions is known to be fairly small \cite{Ho}. 

As one can see from the Fig.1 the corrections are quite significant already at $p_{F}r_e \sim 1$ and grow as the value of $p_F r$
increases.We found $\xi \simeq 0.68$ in the case of neutron matter with scattering length and effective radius being $-18.5$ fm and
$2.7$ fm correspondingly. The large value of the effective radius contribution suggests that one might expect the nonnegligible
contribution form the next term of the low energy expansion of the effective fermion-fermion interaction which is proportional
to $C_4 p^4$. The corresponding T-matrix in a free space takes the form
\begin{equation}
\frac{1}{T_1}=\frac{1}{C_{0}} + \frac{M}{4\pi}\mu  - \frac{C^{'}_2}{C^{2}_{0}}p^2 +  \frac{C^{'}_2}{C^{2}_{0}}\biggl(\frac{C^{'}_2}{C_{0}}
 - \frac{C_4}{C^{'}_{2}}\biggl)p^4  ,  
\end{equation}
where we have written the T-matrix in a form, consistent with the  counting rules suggested in \cite{KSW}.
The coupling $C_4$ entering the effective lagrangian at next-to-next leading order 
can be related to the so called shape parameter \cite{KSW}. Again using the neutron matter parameters
we found approximately $\xi_{C_4} \simeq 0.73$. We note that the size of the correction, while being non-negligible,  
suggests that the contributions of the higher order terms can be neglected.

All that looks reasonable but the word of caution is in order here.
Let us turn to the more general case of the interaction $V_2$ with both energy and momentum dependence. The Feynmann-Galitskii equation
can be solved in the same way and after putting the amplitude $T^{m}_{2}$ on-shell we obtain
\begin{equation}
T^m_2 = T^m_1 - 2 (T^m_1)^2 \frac{C_{2}(\mu)}{C_{0}(\mu)}\frac{M}{6\pi^2}p_F^{3}.
\label{eq:Tm2}
\end{equation}

\begin{figure}
\hspace*{\fill}\includegraphics[height=7cm, width=10cm]{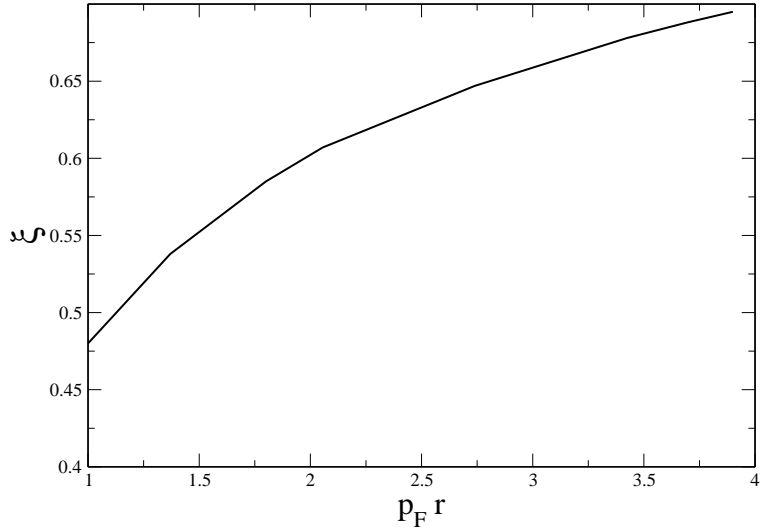}\hspace*{\fill}
\caption{ The  universal parameter $\xi$ as the function of 
 $p_F r$. }
\label{fig:GSE}
\end{figure}

We see that two interactions which resulted in the same physical amplitudes in vacuum, lead to the different on-shell $T$-matrices in the presence
of the fermion medium so that the reparametrisation invariance is not satisfied. Neither is satisfied the RG invariance requirement as
the renormalisation performed at different subtraction points leads to different results for the physical observables. In other words, the physical
 observables still depend on the off-shell behaviour  assumed for the fermion-fermion interaction.
This is clearly the unsatisfactory situation which should be corrected. The general analysis of this problem was given in \cite{Kr2} so that here we give
just a summary of the main points from \cite{Kr2}. Firstly, the hint on how to cancel the unphysical contributions comes from the second term 
in Eq.(17) which is proportional to the density. The same structure arises from a three-body (3B) contact interaction so that the 3B forces could
probably be used to achieve the required cancellation.
Secondly, we note that the contribution of the off-shell term is driven by the 
coupling constant $C_2$ which cannot be extracted from any physical observables which are defined  on-shell. In the lowest order in the off-shell 
coupling $C_2$ the ground state energy
is determined by the Hugenholtz diagrams shown in Fig. 2, 
where the solid dot denotes an in medium $NN$ vertex and thick lines are dressed fermion propagators.  Hugenholtz diagrams are versions of Feynman 
diagrams which explicitly incorporate antisymmetry of the interactions. 
Internal lines represent  Feynman propagators which 
describe both particles and holes. The arrows represent the flow of 
quantum numbers such as baryon number. Each topologically distinct 
diagram should be multiplied by a symmetry factor to take account of the 
number of ways it can be constructed from the antisymmetric vertices. 
More details of these diagrams and the rules for evaluating them can be 
found in the textbooks \cite{BR,NO}. 

The diagrams in 
Fig.~\ref{fig:GSE} give rise to many different contributions, which can be 
identified by iterating the equations for the in-medium NN vertex and 
dressed propagator. The approximations commonly
used in many-body physics, typically amount to replacing the full NN vertex by a
$G$- or $T$-matrix including both particle-particle ($p-p$) and hole-hole ($h-h$) ladders. Including $ph$ rings as well as $pp$ and $hh$
ladders leads to the parquet approximation \cite{BR}.

\begin{figure}
\hspace*{\fill}\includegraphics[height=2.0cm]{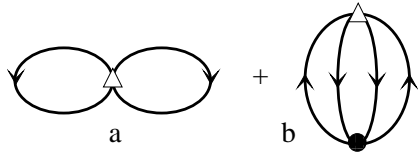}\hspace*{\fill}
\caption{ Hugenholtz diagrams for the ground state energy at first order
order in  $C_2$ (the open triangle).}
\label{fig:GSE}
\end{figure}

As shown in \cite{Kr2} the terms containing the unwanted off-shell contributions can indeed be exactly canceled against the contributions of a
 contact 3B interactions
 with  three distinct topological structures shown in Fig.3.
\begin{figure}
\hspace*{\fill}\includegraphics[height=2.6cm]{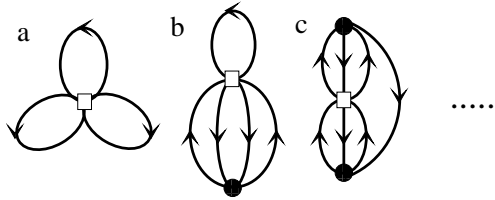}\hspace*{\fill}
\caption{ Hugenholtz diagrams for the ground state energy at first 
order in  the three-body force  (the open square).}
\label{fig:3BF}
\end{figure}
In the simplest Brueckner-Hartree-Fock approximation  
\cite{BR,NO}, in which propagators are dressed and the in-medium NN vertex 
is obtained by iterating the potential in the $p-p$ and $h-h$ channels the unphysical contributions can be shown to cancel with Fig.3(a).
When $p-h$ channel is added so that one iterates the interaction in all ($p-p$, $h-h$ and $p-h$) channels
 all three graphs from Fig.(3) are needed to achieve the required cancellations. 

We note that it is rather hard to make a rigorous statement on whether it is at all possible to formulate the general approach based on the 2B forces 
without the 3B ones and simultaneously satisfying the reparametrisation invariance requirement. The form of the second term in Eq.(17) seems to indicate
 that this is  not possible but this is admittedly suggestive argument. In the context of the above discussion the more rigorous statement would be 
that for the most popular and widely used many-body approaches such as  $p-p$/$h-h$ ladder, parquet and even advanced parquet \cite{BR} approximations
inclusion of the 3B forces is required to satisfy the reparametrisation invariance theorem. An additional support for this statement comes from the EFT
studies of the few-body systems \cite{VK} where the 3B forces are needed to carry out a consistent renormalisation procedure. Moreover, as shown in 
\cite{VK} the corresponding 3B vertex must be promoted to a leading order in effective Lagrangian. 
All that strongly suggests that the 3B forces must necessarily be included at any level of truncations used so far in theoretical calculations. 

The importance of the higher order diagrams with both the 2B and 3B forces could be estimated more quantitatively from some power counting rules.
  Unfortunately, as we  pointed out above, establishing such a counting for 
the strongly interacting Fermi system is still  an open and very challenging problem.

 As we already mentioned, the off-shell parameters cannot be extracted from the on-shell physics so one can 
only theoretically estimate the off-shell contribution and, hence, the strength of the 3B forces needed to cancel it. The possible 
estimate could be based on the assumption
that the term with the coupling $C_2$ gives the contributions of the same order as those related to its  on-shell ``cousin'' $C'_{2}$. It leads to 
\begin{equation}
C_2 \sim C'_{2} \sim  0(1),
\end{equation}
if $\mu \sim p_F$. One notes that the 3B forces will also depend on the subtraction point $\mu$ to satisfy the RG invariance.
Of course, the estimates obtained for the 3B forces are very  crude  and much quantitative work remains to be done to properly take their effect into account.
 Apart from the ``destructive''
role of cancelling the off-shell contributions the 3B forces should also play a ``constructive'' role in bringing the theory to better agreement 
with the data.   The examples include the binding energy of the triton  or low energy neutron-deuteron scattering \cite{Nd}. The other example, 
which is very general 
and hold for any  three body system with infinite scattering length in the  two body subsystems is the Efimov effect \cite{Ef} which states that in such a system 
there exist infinitely many three-body bound states.
It clearly should have a huge influence on the dynamics of the cold Fermi gases near the unitary limit  and should therefore be taken into account
in any consistent theoretical approach. Work in this direction is in progress.

The bottom line of this discussion is that, in spite of the fact that the EFT motivated studies of the Fermi gases near the unitary limit 
may result in the reasonable numbers for the physical observables they should be interpreted with a caution as neither power counting nor 
reparametrisation (and renormalisation) invariance issue is satisfactory implemented in theoretical schemes at present.

\end{document}